\documentclass[english,twocolumn]{revtex4}
\usepackage[T1]{fontenc}
\usepackage[latin9]{inputenc}
\usepackage{textcomp}
\usepackage{bm}
\usepackage{amsmath}
\usepackage{amssymb}
\usepackage{graphicx}
\usepackage{esint}

\makeatletter

\providecommand{\tabularnewline}{\\}

\@ifundefined{textcolor}{}
{%
 \definecolor{BLACK}{gray}{0}
 \definecolor{WHITE}{gray}{1}
 \definecolor{RED}{rgb}{1,0,0}
 \definecolor{GREEN}{rgb}{0,1,0}
 \definecolor{BLUE}{rgb}{0,0,1}
 \definecolor{CYAN}{cmyk}{1,0,0,0}
 \definecolor{MAGENTA}{cmyk}{0,1,0,0}
 \definecolor{YELLOW}{cmyk}{0,0,1,0}
}

\makeatother

\usepackage{babel}
\begin{document}

\title{Absorption Spectra of Astaxanthin Aggregates}

\author{Jan Ol\v{s}ina$^{1}$, Milan Durchan\textsuperscript{2}, Babak
Minofar\textsuperscript{2,3}, Tom\'{a}\v{s} Pol\'{i}vka$^{2}$ and Tom\'{a}\v{s}
Man\v{c}al$^{1}$}

\affiliation{$^{1}$Faculty of Mathematics and Physics, Charles University in
Prague, Ke Karlovu 5, CZ 121 16 Prague 2, Czech Republic}

\affiliation{$^{2}$Faculty of Science, University of South Bohemia, Brani\v{s}ovsk\'{a}
31, 370 05 \v{C}esk\'{e} Bud\v{e}jovice, Czech Republic}

\affiliation{3 Department of Chemistry, Faculty of Science, Kyushu University,
Hakozaki, Higashi-ku, Fukuoka 6-10-1, 812-8581, Japan}
\begin{abstract}
Carotenoids in hydrated polar solvents form aggregates characterized
by dramatic changes in their absorption spectra with respect to monomers.
Here we analyze absorption spectra of aggregates of the carotenoid
astaxanthin in hydrated dimethylsulfoxide. Depending on water content,
two types of aggregates were produced: H-aggregates with absorption
maximum around 390 nm, and J-aggregates with red-shifted absorption
band peaking at wavelengths >550 nm. The large shifts with respect
to absorption maximum of monomeric astaxanthin (470-495 nm depending
on solvent) are caused by excitonic interaction between aggregated
molecules. We applied molecular dynamics simulations to elucidate
structure of astaxanthin dimer in water, and the resulting structure
was used as a basis for calculations of absorption spectra. Absorption
spectra of astaxanthin aggregates in hydrated dimethylsulfoxide were
calculated using molecular exciton model with the resonance interaction
energy between astaxanthin monomers constrained by semi-empirical
quantum chemical calculation. The intramolecular vibrations of astaxanthin
are modeled by a line shape function corresponding to two characteristic
C-C and C=C stretching modes with frequencies of 1150 cm$^{-1}$ and
1520 cm$^{-1}$. The solvent is represented by a spectral density
of harmonic vibrations. The spectral changes at increasing concentrations
of water were assigned to formation of aggregates with decreasing
exposure of the astaxanthin hydrophobic chain to the solvent.
\end{abstract}
\maketitle

\section{Introduction\label{sec:Introduction}}

Carotenoids are a widespread group of natural pigments that attracted
a lot of attention during the past decade due to their rich photophysics
\cite{Polivka2004a-1,Ostroumov-2,Maiuri-3,Polivka-4,Kosumi-5,Gradinaru-6}.
Their photophysical properties are important for their functions in
photosynthetic organisms, where they act as key molecules in both
light-harvesting and photoprotection \cite{Polivka-7,Ruban-8}. Based
on a number of studies of monomeric carotenoids it is now a well-established
fact that the strongly absorbing transition from the ground state
responsible for the characteristic colors of carotenoids is not their
lowest energy transition. Instead, at least one dark singlet state,
denoted as the $S_{1}$ state, lies below the absorbing state which
is denoted as $S_{2}$ state, accordingly \cite{Polivka2004a-1}.
Moreover, recent results indicated that other dark states inaccessible
from the ground state may be located within the $S_{1}$-$S_{2}$
energy gap \cite{Ostroumov-2,Maiuri-3,Polivka-4,Koyama-9}.

While the excited-state properties of monomeric carotenoids were studied
to great details by various experimental and theoretical methods \cite{Polivka2004a-1,Ostroumov-2,Maiuri-3,Polivka-4,Kosumi-5,Gradinaru-6,Polivka-7,Ruban-8,Koyama-9,Starcke-10,Ghosh-11,Kleinschmidt-12},
spectroscopic properties of carotenoid aggregates, which are readily
formed in hydrated polar solvents \cite{Simonyi-13,Billsten-14,Koepsel-15,Mori-16,Ruban-17,Giovannetti-18},
are much less understood. Two types of carotenoid aggregates can be
distinguished according to their absorption spectra. The first type
is associated with a large blue shift of the absorption spectrum of
a monomeric carotenoid that is also accompanied by a loss of vibrational
structure of the $S{}_{2}$ state. These aggregates are usually denoted
as H-aggregates, in which the conjugated chains are closely packed
oriented parallel to each other \cite{Simonyi-13,Spano-19}. The second
aggregation type (J-aggregate) is characterized by a red-shift of
the absorption spectrum, while the resolution of vibrational bands
is mostly preserved. This aggregation (J-type) is likely a result
of a head-to-tail organization of conjugated chains, forming a loose
association of carotenoid molecules \cite{Simonyi-13,Spano-19}. Besides
changes in absorption spectra, carotenoid aggregates also exhibit
significant changes in circular dichroism (CD) \cite{Simonyi-13},
excited-state dynamics \cite{Billsten-14}, and Raman spectra \cite{Wang-20,Wang-21}. 

Recent experimental studies showed that carotenoid aggregates may
play a significant role in various natural and artificial systems.
Carotenoids tend to aggregate when present in lipid bilayers, where
they typically form H-type aggregates \cite{Sujak-22,Gruszecki-23}.
However, it was recently shown that absorption changes consistent
with J-type aggregation may occur in certain carotenoproteins \cite{Aspinall-24,chabera-25}.
In artificial systems, H-aggregates are often formed when carotenoids
are deposited on surfaces. Since assemblies consisting of carotenoid
molecules attached to conducting or semiconducting materials hold
promise to act as photoactive species in dye-sensitized solar cells
\cite{Gao-26,Pan-27,Pan-28,Xiang-29,Wang-30}, understanding the effects
of aggregation on the structure of excited states is an important
factor in controlling the efficiency of such devices. Also, aggregation
of carotenoids facilitates a homofission process \cite{Smith-31},
in which the $S_{1}$ state of carotenoid aggregate breaks down to
form two triplets \cite{Wang-20,Wang-21}. When attached to an electron
acceptor (e.g. TiO$_{2}$ semiconductor), this process promises to
generate two electrons from a single absorbed photon, thus it is extremely
attractive for the future design of devices aiming for conversion
of solar energy to electricity. 

Carotenoid astaxanthin, which is the subject of this study, is known
mainly as the colorant of the lobster carapace where it is the only
pigment of the carotenoprotein crustacyanin \cite{Cianci-32}. Upon
binding to crustacyanin, absorption spectrum of astaxanthin undergoes
a large red shift whose origin is still a matter of debate and excitonic
interaction due to aggregation has been one of the suggested mechanisms
of the red shift \cite{Wijk-33}. A few experimental studies of astaxanthin
aggregation in hydrated solvents demonstrated that at ambient temperatures
astaxanthin predominantly forms H-aggregates in hydrated acetone \cite{Mori-16}
or methanol \cite{Koepsel-15,Giovannetti-18}. J-aggregates are generated
and stabilized only upon heating up the system above 30 \textdegree{}C
\cite{Giovannetti-18}. Yet, no theoretical studies aiming for explanation
of aggregation-induced changes of absorption spectra of astaxanthin
were carried out so far. 

In this work, we explain absorption spectra of astaxanthin aggregates
employing Frenkel exciton model for resonance interaction within astaxanthin
aggregates, constrained by molecular dynamics simulations of an astaxanthin
dimer. Two related models of astaxanthin aggregates are developed.
For the purpose of calculating absorption spectra of astaxanthins,
we introduce a \emph{molecular exciton model}. In order to estimate
one of the essential parameters of the exciton model, the resonance
coupling, and in order to constrain its values for subsequent fitting,
we use a more detailed, but still simplified, semi-empirical $\pi$-electron
\emph{molecular orbital model} of the astaxanthin. Excitonic model
with explicit treatment of fast vibrational modes of carotenoids was
used by Spano \emph{et al.} \cite{Spano-19} to explain absorption
and CD spectra of carotenoids. Here, the general approach is similar,
although we do not consider the vibrational modes explicitly, but
describe them by energy gap correlation function.

In Section \ref{sec:Preparation-of-astaxanthin} we discuss the preparation
of the astaxanthin aggregates in different solutions. Section \ref{sec:Absorption-Spectrum},
presents the lineshape method of calculation of absorption spectrum.
The intramolecular vibrations of the astaxanthin are modeled by a
two modes with frequencies $\omega_{vib}^{(1)}=1150$ cm$^{-1}$ and
$\omega_{vib}^{(2)}=1520$ cm$^{-1}$ and the solvent is represented
by a spectral density of harmonic vibrations (e.g. instantaneous normal
modes). In Section \ref{sec:Excitonic-Model} we discuss the molecular
exciton model, where the astaxanthin is represented by two electronic
levels, (the ground state $|g\rangle$ and exited state $|e\rangle$)
standing for the $S_{0}$ and the allowed $S_{2}$ states of the astaxanthin,
respectively. We also discuss calculation of the absorption spectrum
of the aggregates of astaxanthins based on such a model. The crucial
parameter, which characterizes the aggregates assembled after the
water is added into the solution, is the resonance coupling between
the members of the aggregate. We perform molecular dynamics simulations
in Section \ref{sec:Molecular-Dynamics-Simulation} to obtain an insight
into the structure of the astaxanthin dimer in water. The average
distance between the molecules in this case is used to qualitatively
estimate the resonance coupling in Section \ref{sec:Resonance-Coupling}.
In Section \ref{sec:Spectra-of-Aggregate} we compare the spectra
of different aggregates of astaxanthin and fit the experimental absorption
spectra obtained under various solution conditions.

\section{Preparation of astaxanthin aggregates\label{sec:Preparation-of-astaxanthin}}

\begin{figure}
\includegraphics[width=1\columnwidth]{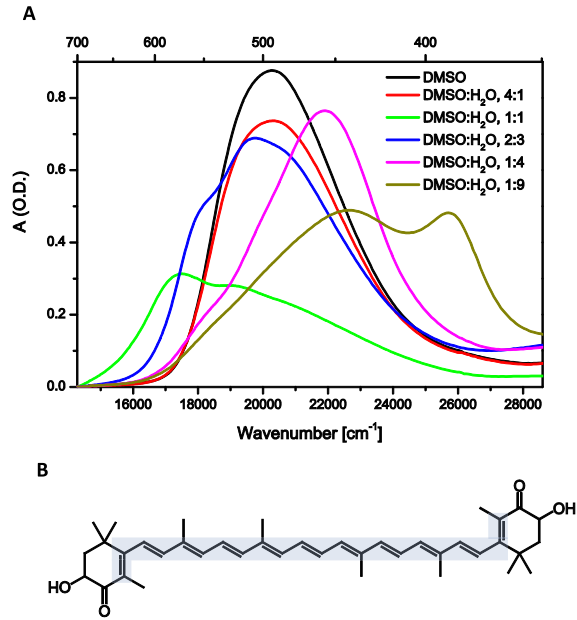}

\caption{\label{fig:TPo1}Part A: Absorption spectra of astaxanthin in pure
DMSO and in various DMSO/water mixtures. In all samples, the concentration
of astaxanthin is 20 (micro)M. Part B: Molecular structure of astaxanthin
with the part of the $\pi-$conjugated system used for calculations
marked in light blue. }
\end{figure}

Absorption spectra of astaxanthin monomers and aggregates in hydrated
dimethylsulfoxide (DMSO) along with molecular structure of astaxanthin,
are shown in Fig.~\ref{fig:TPo1}. DMSO was chosen due to good solubility
of astaxanthin in this solvent and also due to the large viscosity
of DMSO which stabilizes aggregates. Also, as descried below, both
types of aggregates can be readily formed in DMSO. Absorption spectrum
of astaxanthin monomer in DMSO peaks at 20280 cm$^{-1}$ (493 nm).
The $S_{0}$-$S_{2}$ transition of astaxanthin monomer does not exhibit
any vibrational structure, a phenomenon characteristic of carotenoids
having conjugated carbonyl groups \cite{Frank-34,Zigmantas-35}. Addition
of water into the stock solutions of astaxanthin in DMSO changes absorption
spectra significantly. All astaxanthin aggregates were prepared in
a way that astaxanthin concentration, 20 $\mu$M, remains constant
in all solvent/water mixtures. 

In hydrated DMSO both J- and H-types of astaxanthin aggregates are
readily formed. At DMSO:water ratio of 1:1, a clear band 17450 cm$^{-1}$
(573 nm) indicates that most of the astaxanthin molecules in solution
formed J-aggregates. At increasing water concentration, J-aggregates
transforms to H-aggregates as indicated by narrow absorption band
at 25640 cm$^{-1}$ (390 nm). Thus, DMSO promotes formation of both
types of aggregates with J-type generated only in rather narrow water
concentration in agreement with earlier reports on astaxanthin \cite{Giovannetti-18}
or zeaxanthin \cite{Billsten-14} in methanol. However, while in previous
studies using methanol or acetone required higher initial concentrations
of carotenoid to induce stable J-aggregates, in DMSO the J-aggregates
are generated even at moderate concentration of 20 $\mu$M. Since
we kept the concentration of astaxanthin constant during experiments,
absorption spectra shown in Fig.~\ref{fig:TPo1} also demonstrate
how the extinction coefficient of the $S_{0}$-$S_{2}$ transition
changes upon aggregation. In both solvents, formation of aggregates
is accompanied by significant decrease of extinction coefficient of
both J- and H-aggregates.

\section{Calculation of Absorption Spectrum\label{sec:Absorption-Spectrum}}

In order to characterize absorption spectra of astaxanthin aggregates,
we apply a line shape model based on second cumulant expansion of
the electron-phonon coupling. In contrast to recently reported vibronic
exciton approach aiming to explain absorption and CD spectra of aggregates
of another carotenoid, lutein \cite{Spano-19}, we do not treat vibrational
modes of the carotenoid explicitly. In the case of a carotenoid monomer,
the two approaches would lead to the same results (second cumulant
expansion is exact for calculation of absorption spectra of harmonic
systems \cite{Doll2008a}). The reduction to a purely electronic exciton
approach, which is very successful in describing chlorophyll based
aggregates \cite{Cho2005a}, simplifies the treatment of the system
significantly. It is also motivated by the lack of detailed vibrational
structure in astaxanthin and by only a small perturbation of the vibrational
band structure upon aggregation reported for other carotenoids \cite{Wang2012} 

The astaxanthin molecule in solution can be represented by the following
Hamiltonian
\begin{equation}
H_{{\rm car}}=(\epsilon_{g}+T+V_{g})|g\rangle\langle g|+(\epsilon_{e}+T+V_{e})|e\rangle\langle e|,\label{eq:HamCar}
\end{equation}
where the states $|g\rangle$ and $|e\rangle$ were introduced in
Section \ref{sec:Introduction}, $\epsilon_{g}$ and $\epsilon_{e}$
are the ground and excited state electronic energies, $T$ represents
the kinetic energy operator of the intramolecular nuclear modes of
the carotenoid and the kinetic energy of the solvent DOF, and $V_{g}$
and $V_{e}$ represent the potential energy surfaces of these DOF
in the electronic ground and excited states, respectively. The Hamiltonian,
Eq.~(\ref{eq:HamCar}), can be split into the usual system (S), bath
(B) and interaction (I) parts as follows
\[
H_{{\rm car}}=\underbrace{T+V_{g}}_{H_{B}}+\underbrace{\epsilon_{g}|g\rangle\langle g|+[\epsilon_{e}+\langle V_{e}-V_{g}\rangle_{eq}]|e\rangle\langle e|}_{H_{S}}
\]
\begin{equation}
+\underbrace{\Delta V|e\rangle\langle e|}_{H_{I}}.
\end{equation}
In the interaction Hamiltonian $H_{I}$ we introduced so-called energy
gap operator 
\begin{equation}
\Delta V=V_{e}-V_{g}-\langle V_{e}-V_{g}\rangle_{eq},
\end{equation}
where $\langle V_{e}-V_{g}\rangle_{{\rm eq}}={\rm tr}\{(V_{e}-V_{g})W_{{\rm eq}}\}$
is the equilibrium expectation value of the difference between the
excited state and the ground state nuclear PES. 

To describe spectroscopy, let us introduce the light-matter interaction
Hamiltonian in the semiclassical dipole approximation, i.e. 
\begin{equation}
H_{E}(t)=-\bm{\mu}\cdot\bm{E}(t),
\end{equation}
where the dipole moment operator reads
\begin{equation}
\bm{\mu}=\bm{d}|g\rangle\langle e|+\bm{d}^{*}|e\rangle\langle g|,
\end{equation}
with $\bm{d}$ independent of the solvent and intramolecular carotenoid
DOF (Condon approximation). The absorption coefficient can then be
expressed in terms of linear response function $S^{(1)}(t)$ as \cite{MukamelBook}
\begin{equation}
\alpha(\omega)=\frac{\omega}{n(\omega)}{\rm Re}\int\limits _{0}^{\infty}d\omega\ S^{(1)}(t)e^{i\omega t}.\label{eq:abs1}
\end{equation}
For a two-level system interacting with harmonic oscillators, exact
expression for the linear response function can written down in terms
of the so-called \emph{line shape function} 
\begin{equation}
g(t)=\frac{1}{\hbar^{2}}\int\limits _{0}^{t}d\tau\int\limits _{0}^{\tau}d\tau^{\prime}C(\tau^{\prime}),
\end{equation}
where $C(t)$ is so-called bath or \emph{energy gap correlation function}
\begin{equation}
C(t)={\rm tr}\{e^{\frac{i}{\hbar}H_{B}t}\Delta Ve^{-\frac{i}{\hbar}H_{B}t}\Delta V\ W_{{\rm eq}}\}.\label{eq:corrfce}
\end{equation}
Here, $W_{{\rm eq}}$ is the equilibrium density matrix of the intramolecular
and solvent DOF. The first order response now reads as 
\begin{equation}
S^{(1)}(t)=\frac{i}{3}\Theta(t)|d|^{2}e^{-g(t)-i\omega_{eg}t}.
\end{equation}
The prefactor $\frac{1}{3}$ results from the averaging of the response
function over an isotropic distribution of the dipole moment vectors
of the molecule in space. By prescribing the correlation function
$C(t)$ in a form of two independent contributions 
\begin{equation}
C(t)=C_{{\rm vib}}(t)+C_{{\rm solvent}}(t),
\end{equation}
where the contribution of intramolecular vibrations 
\[
C_{{\rm vib}}(t)=\lambda_{{\rm vib}}\Big(\frac{2k_{B}T}{\hbar}\cos(\omega_{{\rm vib}}t)
\]
\begin{equation}
-i\omega_{{\rm vib}}\sin(\omega_{{\rm vib}}t)\Big)e^{-\Gamma_{{\rm vib}}t/2},\label{eq:Cvib}
\end{equation}
and the contribution of solvent reads as 
\begin{equation}
C_{{\rm solvent}}(t)=\lambda_{{\rm solv}}\left(\frac{2k_{B}T}{\hbar}-i\Gamma\right)e^{-\Gamma t},\label{eq:Cbath}
\end{equation}
we completely specify the absorption spectrum of the monomer. Details
of the model correlation functions Eqs.~(\ref{eq:Cvib}) and (\ref{eq:Cbath})
can be found in the Ref. \cite{MukamelBook}.

The correlation function, Eq.~(\ref{eq:corrfce}), enables us to describe
another important contribution to the absorption spectrum, so-called
\emph{static disorder} or \emph{inhomogeneous broadening}. Provided
that there is a certain distribution of transition energies $\hbar\omega_{eg}$
such that it is characterized by a constant energy gap correlation
function $C_{{\rm dis.}}(t)=\Delta^{2}$, this could be easily incorporated
into the absorption spectrum formula, Eq.~(\ref{eq:abs1}), which
now reads
\begin{equation}
\alpha(\omega)=\frac{\omega|d|^{2}}{3n(\omega)}{\rm Im}\int\limits _{0}^{\infty}d\omega\ e^{-g(t)-\frac{\Delta^{2}}{2\hbar^{2}}t^{2}-i(\omega_{eg}-\omega)t}.\label{eq:abs2}
\end{equation}

Instead of Eq.~(\ref{eq:abs2}), one can alternatively use explicit
drawing of the energy gap value out of a Gaussian distribution with
the width $\Delta^{2}$. For a monomer, both approaches are equivalent.
For aggregate, the line-shape function depends on the aggregate structure,
and the simulation of disorder has to be preformed explicitly.

\begin{figure}
\includegraphics[width=1\columnwidth]{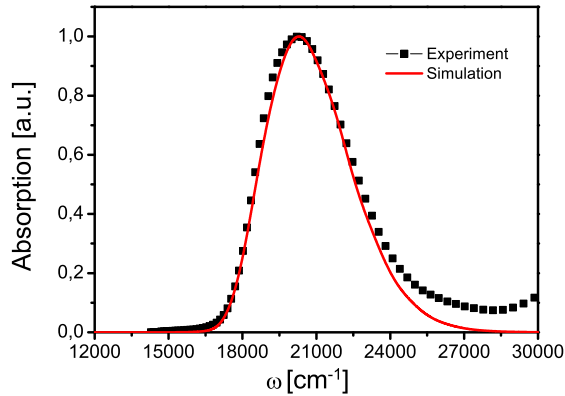}

\caption{\label{fig:Monomer}Experimental (red) and calculated (black) absorption
spectrum of a monomer astaxanthin in DMSO. The parameters of calculation
are: transition frequency $\omega_{eg}=20700$ cm$^{-1}$, vibrational
frequencies $\omega_{{\rm vib}}^{(1)}=1520$ cm$^{-1}$, and $\omega_{{\rm vib}}^{(2)}=1150$
cm$^{-1}$, reorganization energies of the vibrations $\lambda_{{\rm vib}}^{(1)}=1220$
cm$^{-1}$, and $\lambda_{{\rm vib}}^{(2)}=250$ cm$^{-1}$, damping
time of the vibrations $\tau_{{\rm vib}}^{(1,2)}=\frac{1}{\Gamma_{{\rm vib}}}=1$
ps, solvent reorganization energy $\lambda_{{\rm solv}}=550$ cm$^{-1}$,
solvent correlation time $\tau_{c}=100$ fs. }
\end{figure}

\section{Excitonic Model\label{sec:Excitonic-Model}}

When aggregates are formed in the solution, astaxanthin molecules
start to interact with each other. We will use the subscript $n$
and the superscript $(n)$ to distinguish individual molecules of
the aggregate in this section. The total Hamiltonian reads
\begin{equation}
H=\sum_{n=1}^{N}H_{{\rm car}}^{(n)}+H_{{\rm int}},
\end{equation}
where $H_{{\rm int}}$ represents all possible interaction terms that
follow from quantum chemistry. Electronic states of the aggregates
can be expressed using the basis composed of excitations of individual
monomers. The aggregate is in the ground state when all the molecules
in the aggregate are in the ground state. Thus we can write
\begin{equation}
|G\rangle\equiv\prod_{n=1}^{N}|g_{n}\rangle,
\end{equation}
where $|g_{n}\rangle$ is the ground state of the n$^{th}$ astaxanthin.
The excited states of the aggregate in the spectral region around
the $S_{2}$ band of the monomeric astaxanthin are formed from single
excitation states 
\begin{equation}
|E_{k}\rangle=\prod_{n=1}^{k-1}|g_{n}\rangle|e_{k}\rangle\prod_{m=k+1}^{N}|g_{m}\rangle,
\end{equation}
where $|e_{k}\rangle$ is the excited $S_{2}$ state of the k$^{th}$
astaxanthin in the aggregate. 

Expressed in these states, the aggregate Hamiltonian reads
\[
H=H_{B}+\sum_{n}(\tilde{\epsilon}_{n}+\Delta V^{(n)})|E_{n}\rangle\langle E_{n}|
\]
\begin{equation}
+\sum_{m\neq n}J_{mn}|E_{m}\rangle\langle E_{n}|,\label{eq:totHam}
\end{equation}
where the ground state energy of the aggregate $\epsilon_{g}$ is
set to zero, $\tilde{\epsilon}_{n}=\hbar\omega_{eg}^{(n)}$ is the
excited state energy of the n$^{th}$ astaxanthin ($\tilde{\epsilon}_{n}=\epsilon_{e}+{\rm tr}_{{\rm bath}}\{(V_{e}^{(n)}-V_{g}^{(n)})W_{eq}\}$)
and $J_{nm}$ is the electrostatic coupling between two astaxanthins
\begin{equation}
J_{nm}={\rm tr}_{{\rm bath}}\{\langle E_{n}|H_{int}|E_{m}\rangle W_{eq}\}.
\end{equation}

The electronic eigenstates of the Hamiltonian, Eq.~(\ref{eq:totHam}),
can be found by the diagonalization of its purely electronic part
(i.e. ignoring $\Delta V^{(n)}$ and $H_{B}$ in Eq.~(\ref{eq:totHam})).
The new excited eigenstates and their energies will be denoted $|\bar{E}_{n}\rangle$
and $\bar{\epsilon}_{n}$, respectively. The total Hamiltonian now
reads
\[
H=H_{B}+\sum_{n}(\bar{\epsilon}_{n}+\Delta\bar{V}_{nn})|\bar{E}_{n}\rangle\langle\bar{E}_{n}|
\]
\begin{equation}
+\sum_{m\neq n}\Delta\bar{V}_{mn}|\bar{E}_{m}\rangle\langle\bar{E}_{n}|.\label{eq:totHam2}
\end{equation}
The last term represents energy transfer between the electronic eigenstates
of the aggregate, the effects of which lead to broadening of the absorption
spectra. We will ignore these effects in further discussion and assume
them to be smaller than the effects of broadening due to pure dephasing
(the diagonal term $\Delta\bar{V}_{nn}$) and the disorder. The pure
dephasing terms read as
\begin{equation}
\Delta\bar{V}_{nn}=\sum_{k}|\langle\bar{E}_{n}|E_{k}\rangle|^{2}\Delta V^{(k)}.
\end{equation}
 The delocalized eigenstates $|\bar{E}_{n}\rangle$ are usually termed\emph{
molecular excitons} \cite{ValkunasBook}. Thus we arrive at a situation
which is similar to calculation of the absorption spectrum of the
monomer, Eq.~(\ref{eq:abs2}), except of the number of involved states
which is equal to the number of members of the aggregate. We thus
have 
\begin{equation}
\alpha(\omega)=\frac{\omega}{3n(\omega)}\langle\sum_{n}|d_{n}|^{2}{\rm Im}\int\limits _{0}^{\infty}d\omega\ e^{-g_{nn}(t)-i(\omega_{ng}-\omega)t}\rangle_{diss},
\end{equation}
where $\langle\dots\rangle_{diss}$ denotes explicit averaging over
energy gap disorder, $d_{n}$ are the transition dipole moments of
the $n^{th}$ eigenstate, and $g_{nn}(t)$ is the lineshape function
corresponding to the energy gap operator $\Delta\bar{V}_{nn}$ in
the Hamiltonian Eq.~(\ref{eq:totHam2}). Assuming that the bath fluctuations
on different molecules are the same, but are not correlated, we can
write for the line shape functions
\begin{equation}
g_{nn}(t)=\sum_{k}|\langle\bar{E}_{n}|E_{k}\rangle|^{4}g(t),
\end{equation}
where $g(t)$ is the lineshape function of a monomeric molecule.

\section{Molecular Dynamics Simulation\label{sec:Molecular-Dynamics-Simulation}}

\begin{figure}
\includegraphics[width=1\columnwidth]{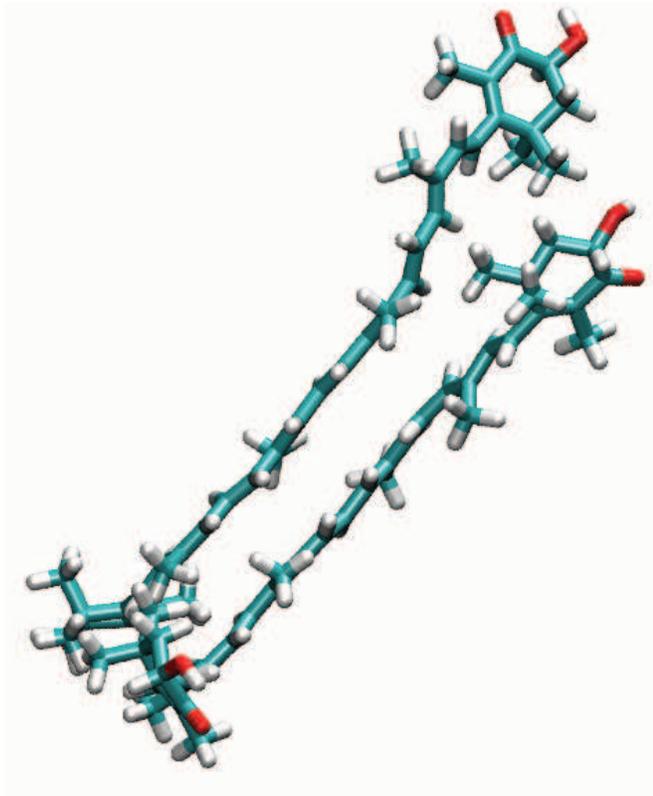}

\caption{\label{fig:TPo2}Structure of astaxanthin dimer resulting from molecular
dynamics simulations. Carbon atoms are shown in blue, hydrogens are
white and oxygens red.}
\end{figure}

In order to obtain parameters for calculation of excitonic interaction,
aggregation of astaxanthin dimer in aqueous solution was studied by
means of classical molecular dynamics (MD) simulations with non-polarizable
force fields. The simulated system consisted of two astaxanthin molecules
and 2700 water molecules forming a slab of liquid in the center of
the simulation box, with two water/vapor interfaces perpendicular
to the z axis. The x, y and z-dimensions of the rectangular simulation
box were set to 50.0, 50.0 and 100.0 Å, respectively with periodic
boundary conditions in all three dimensions. The astaxanthin carotenoid
molecule was modeled by using the general Amber force field (GAFF)
\cite{Wang-36} and water molecules were modeled by using SPC/E model
\cite{Berendsen-37}. To obtain partial charges of astaxanthin molecule,
\emph{ab initio} geometry optimization of single astaxanthin molecule
was performed using the Gaussian 03 package \cite{gaussian03-38}
by employing the density functional theory B3LYP/cc-pVT. After the
optimization by \emph{ab initio} calculations, restricted electrostatic
potential (RESP) procedure was applied using Antechamber program \cite{Wang-39},
which is implemented in the Amber program package \cite{amber8-40}
for obtaining the atomic partial charges. For preparation of the initial
configuration, two astaxanthin molecules with distance of larger than
10 Å were solvated in water box, which was followed by energy minimization
for a few thousands steps. The system was then equilibrated for 2
ns followed by a 5 ns production run. The simulation was carried out
in the NVT ensemble at 300 K, and temperature was controlled by the
Berendsen thermostat \cite{Berendsen-41}. Equations of motion were
integrated using the Leapfrog algorithm with the time step of 1 fs.
The non-bonding interactions were cut at distance of 12.0 Å, and long-range
electrostatic interactions were treated by the Particle Mesh Ewald
procedure \cite{Daren-42,Essmann-43}. All bonds involving hydrogen
atoms were constrained using the SHAKE algorithm \cite{Ryckaert-44}
and a trajectory was constructed by sampling the coordinates at each
5 ps for further analysis such as distance between conjugated chains
and head groups. All MD simulations were performed by employing Amber85
program package, and VMD program was used for visualizations and preparation
of snapshots and distances \cite{Humphrey-45}. Fig.~\ref{fig:TPo2}
shows typical snapshot from MD simulation which shows the aggregation
taking place when astaxanthin molecules solvate in the aqueous solutions.
The structure of astaxanthin dimer is stabilized within first few
ps, which was revealed by analyzing the trajectory from MD simulation.
The conjugated chains of two astaxanthin molecules have in average
the distance of 4.1 Å though the head groups can have even closer
distances during the MD simulation. 

\section{Resonance Coupling between Carotenoid Molecules\label{sec:Resonance-Coupling}}

Within the usual excitonic model one could calculate resonance coupling
between two molecules in dipole-dipole approximation. Transition dipole
moment vector and mutual positions of the molecules are parameters
of the model. This approximation is applicable to well spaced compact
molecules, such as chlorophylls in photosynthetic antennae (i.e. in
proteins), but fails for chain-like molecules such as astaxanthins
in solution. Obviously, the distance between two astaxanthins can
be much smaller than the extent of the molecule, in which case dipole-dipole
approximation breaks down. In order to estimate the coupling between
astaxanthins, we need to use quantum chemical packages, or, as we
do here, to construct a phenomenological model carotenoid with a qualitative
description of the molecular states. We aim at a qualitative theory
of coupling, which allows us to make claims beyond the validity of
the dipole-dipole approximation.

The $S_{2}$ state of a carotenoid originates from the excitation
of the conjugated $\pi$-electronic system. Therefore, we consider
only the delocalized $\pi$-electrons, and we calculate its wavefunction
using an H\"{u}ckel type semi-empirical method. Our model carotenoid
is made of carbon atoms with bond length $1.4$ Å and with positions
chosen in such way, that all bonding angles are 120\textdegree{}.
The structure is depicted on Fig.~\ref{fig:TPo1}. We describe the
$\pi$-electron p-orbital wavefunctions by the STO-3G basis of atomic
orbitals \cite{Szabo1996a}. 

Inspired by the extended H\"{u}ckel method (EHM) \cite{Hoffmann1963},
we define the H\"{u}ckel Hamiltonian in the form
\begin{equation}
\langle\phi_{\mu}|H^{Huckel}|\phi_{\nu}\rangle=H_{\mu\nu}^{Huckel}=\delta_{\mu\nu}\alpha+(S_{\mu\nu}-\delta_{\mu\nu})\beta,\label{eq:HuckelHam}
\end{equation}
where $S_{\mu\nu}\equiv\langle\phi_{\mu}|\phi_{\nu}\rangle$ are overlap
integrals of the atomic orbitals $|\phi_{\mu}\rangle$ and $\alpha$,
$\beta$ are parameters of the method constructed from valence state
ionization potentials \cite{Hoffmann1963}. By solving eigenvalue
problem in a form
\begin{equation}
\sum_{\nu}H_{\mu\nu}^{Huckel}c_{\nu i}=\varepsilon_{i}\sum_{\nu}S_{\mu\nu}\, c_{\nu i},\;\label{eq:HuckelEqnStart}
\end{equation}
we obtain the expansion coefficients $c_{i}$ of the molecular orbitals
in the basis of the atomic orbitals. Eq.~(\ref{eq:HuckelEqnStart})
can be rewritten as
\begin{equation}
\alpha^{\prime}\sum_{\nu}\delta_{\mu\nu}c_{\nu i}=\varepsilon{}_{i}^{\prime}\sum_{\nu}S_{\mu\nu}c_{\nu i}
\end{equation}
with $\alpha^{\prime}=\alpha-\beta$ and $\varepsilon{}_{i}^{\prime}=\varepsilon{}_{i}-\beta$.
We fit the parameter $\alpha'$ to reproduce values of HOMO-LUMO transition
energies of several polyenes of a different length reported in Ref.
\cite{Smith2004}. Because we are interested in only in the energy
difference, the particular value of the parameter $\beta$ is irrelevant,
and we may work with just $\alpha^{\prime}$. Unlike in the original
EHM, this value does not have a direct physical interpretation. The
comparison with polyens yields
\begin{equation}
\alpha^{\prime}=66.48\;\mathrm{eV}.
\end{equation}
Using the obtained molecular orbitals, we calculate the HOMO-LUMO
energy gap and the related transition dipole moment in a standard
manner. We get 
\begin{equation}
\begin{array}{c}
\Delta E^{HOMO-LUMO}=20742\;{\rm cm}^{-1},\\
|\mathbf{d}_{trans.}^{HOMO-LUMO}|=28.6\;\mathrm{D\;.}
\end{array}
\end{equation}

In order to calculate the resonance coupling between two astaxanthins,
we evaluate the corresponding matrix element of the electrostatic
interaction Hamiltonian \cite{Madjet2006} 
\begin{align}
 & V_{ab,a'b'}=\int d\mathbf{r}_{1}\dots d\mathbf{r}_{N}\int d\bar{\mathbf{r}}_{1}\dots d\bar{\mathbf{r}}_{N}\nonumber \\
 & \Psi_{a}^{*}(\mathbf{r}_{1},\dots,\mathbf{r}_{N})\Psi_{b}^{*}(\mathbf{\bar{r}}_{1},\dots,\mathbf{\bar{r}}_{N})\nonumber \\
 & \times\Bigg\{\sum_{i,j}\frac{1}{|\mathbf{r}_{i}-\mathbf{\bar{r}}_{j}|}-\sum_{i,J}\frac{Z_{J}}{|\mathbf{r}_{i}-\mathbf{\bar{R}}_{J}|}-\sum_{I,j}\frac{Z_{I}}{|\mathbf{R}_{I}-\mathbf{\bar{r}}_{j}|}\nonumber \\
 & +\sum_{I,J}\frac{Z_{I}Z_{J}}{|\mathbf{R}_{I}-\mathbf{\bar{R}}_{J}|}\Bigg\}\Psi_{a'}(\mathbf{r}_{1},\dots,\mathbf{r}_{N})\Psi_{b'}(\mathbf{\bar{r}}_{1},\dots,\mathbf{\bar{r}}_{N})\;.\label{eq:Resonance coupling Hamiltonian}
\end{align}
Here, $N$ is the number of all electrons of given molecule, $Z_{I}$
are the nuclear charges in units of the elementary charge, $\mathbf{r}_{i}$
and $\mathbf{\bar{r}}_{i}$ are coordinates of electrons of astaxanthin
1 and 2, $\mathbf{R}_{I}$ and $\mathbf{\bar{R}}_{I}$ are coordinates
of nuclei of astaxanthin 1 and 2 and $\Psi_{a}(\mathbf{r}_{1},\dots,\mathbf{r}_{N})$,
$\Psi_{b}(\mathbf{\bar{r}}_{1},\dots,\mathbf{\bar{r}}_{N})$ are complete
electronic wavefunctions of astaxanthin 1 and astaxanthin 2 respectively.
They are constructed as Slater determinants from atomic orbitals.
Overlaps between atomic orbitals of different astaxanthins are neglected.
Contributions of particular terms of Hamiltonian, Eq.~(\ref{eq:Resonance coupling Hamiltonian}),
can be found by using the Slater-Condon rules for matrix elements
\cite{Szabo1996a}.

\section{Absorption Spectra of Astaxanthin Aggregates \label{sec:Spectra-of-Aggregate}}

\begin{figure}[h]
\includegraphics[width=0.9\columnwidth]{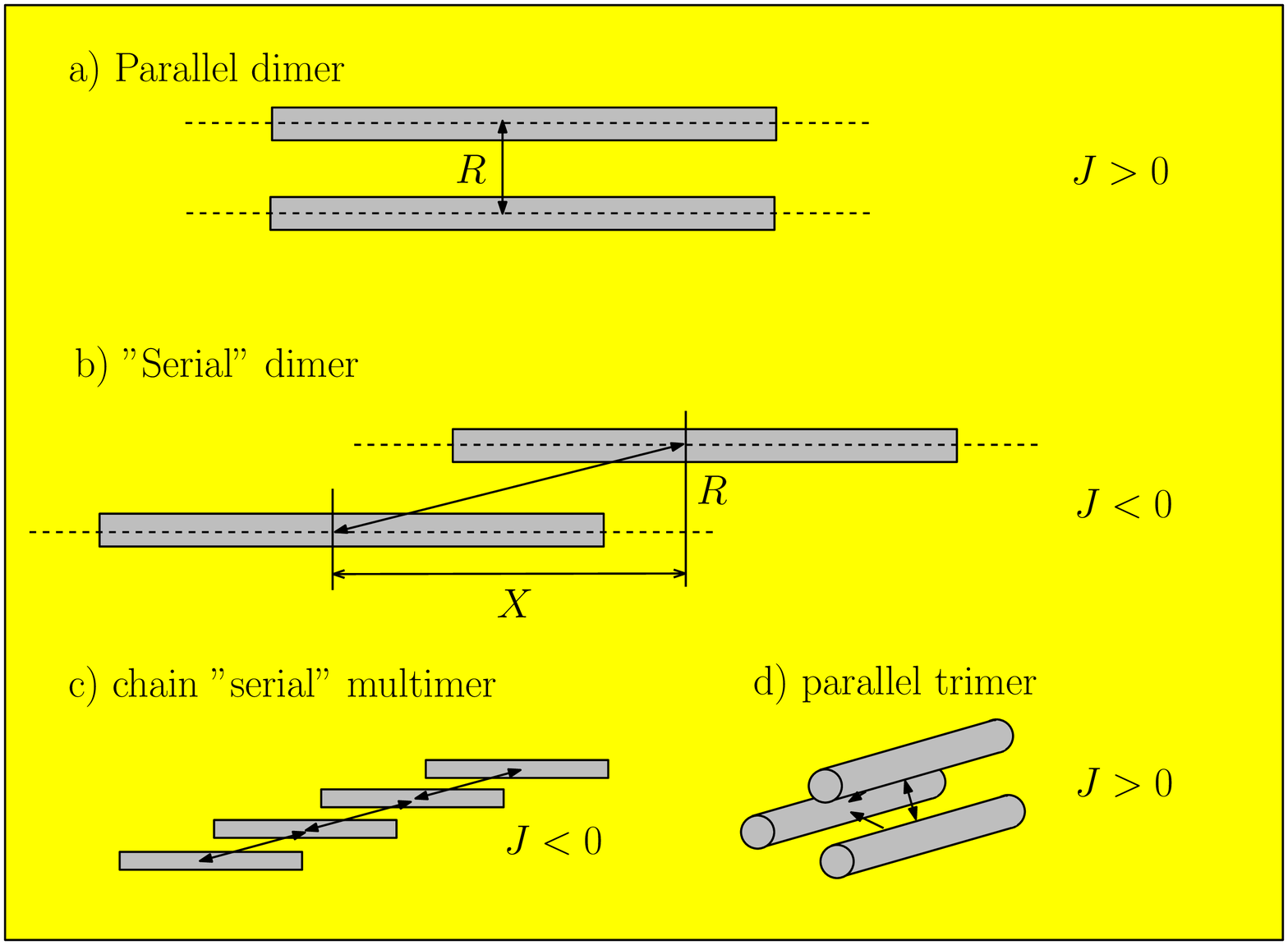}

\caption{\label{fig:Characteristic-structures-of}Characteristic structures
of the astaxanthin dimers and aggregates used for calculation of absorption
spectra. Part a) Parallel dimer and similar slightly shifted arrangements
are characterized by $J>0$. Part b) The ``serial'' dimer with large
shift $X$ is characterized by $J<0$. Part c) Larger aggregates of
the ``serial'' character can have a form of chains. Part d) The
parallel dimer can form stacks or more compact structures such as
a trimer with equal distance between the conjugated chains of the
three molecules. }
\end{figure}

Molecular dynamics simulations show that in water carotenoids tend
to minimize the exposure of their hydrophobic chain to the solvent.
Thus a parallel dimer (see Figs.~\ref{fig:TPo2} and \ref{fig:Characteristic-structures-of})
is a possible arrangement. The dipole-dipole coupling formula predicts
a positive value $J$ of the resonance coupling. Excitonic model predicts
in such a dimer existence of two exited states, which have a form
of a symmetric and anti-symmetric linear combinations of the $S_{2}$
excited states of individual astaxanthins. For the parallel dimer
arrangement, optical excitation of the energetically higher state
is allowed, whereas the lower lying state is dark. Thus one expects
a blue shift of the absorption spectrum upon aggregation. 

\begin{figure}[h]
\includegraphics[width=0.9\columnwidth]{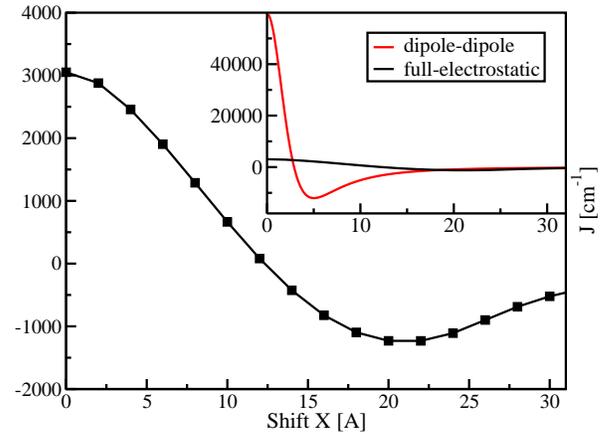}

\caption{\label{fig:Dependence-of-the}Dependence of the resonance coupling
on the mutual shift $X$ (see Fig.~\ref{fig:Characteristic-structures-of})
of the astaxanthin molecules . }
\end{figure}

However, in a mixture of DMSO and water, at the percentage of water
of $50-60$ we observe a red shift (see Fig.~1). Such a shift requires
negative value of the resonance coupling. Dipole-dipole coupling formula
results in negative coupling for a ``serial'' dimer (see Fig.~\ref{fig:Characteristic-structures-of})
where the centers of the astaxanthins are shifted by a sufficient
distance $X$. As we discussed above, dipole-dipole coupling is not
a reliable approximation for astaxanthins, and we therefore performed
a calculation for different values of $X=0,\dots,30$ Å and $R=4.1$
Å (the value predicted by MD simulations in Section \ref{sec:Molecular-Dynamics-Simulation})
with our semi-empirical method. The resulting values of coupling are
presented in Fig.~\ref{fig:Dependence-of-the} and the comparison
of the two predictions is made in the inset. Both results coincide
for large distances $R$ and $X$ where dipole approximation holds.
One can immediately notice that dipole-dipole formula grossly overestimates
the positive coupling for $X=0$, it predicts a rapid switch to negative
values of $J$ (at $X=2.9$ Å) and even for larger $X$ it overestimates
the magnitude of the negative coupling. Our H\"{u}ckel-type calculation,
on the other hand, predicts moderate values of the resonance coupling
(between $-1500$ and $3000$ cm$^{-1}$), and the switch from positive
to negative values occuring at larger $X$, roughly when $X$ is comparable
with the half length of the astaxanthin conjugated chain. Since our
calculations of the resonance coupling are too simple to aim at quantitative
results, we will use the parameter $J$ for selected structures as
a fitting parameter in the next section. We also constrain the resonance
coupling to values up to few thousands of cm$^{-1}$.

To analyze the absorption spectra of astaxanthin in DMSO with varying
percentage of water, we calculated spectra of the two representative
mutual configurations of astaxanthin molecules, namely those with
positive and negative resonance coupling energy $J$. We start with
a simple idea that with increasing percentage of water in the solutions,
it will become energetically advantageous for the molecules to minimize
the surface of their hydrophobic chains exposed to water. With small
amount of water, the free energy profile will be rather flat and we
might preferentially find the molecules in the arrangement similar
to that one of the ``serial'' dimer (see Fig.~\ref{fig:Characteristic-structures-of}).
With increasing amount of water a strict parallel dimer arrangement
will be more and more preferred. 

It is in no way plausible that one could reproduce the absorption
spectrum by composing it from a small number discrete structures,
e.g. dimers and trimers of particular value of coupling $J$. It is
more likely that there exists a smooth distribution of structures
depending on the particular interactions between molecules and the
solvent at a given concentration of water. This would result in a
distribution of values of the coupling $J$. A direct simulation of
the disordered structure of aggregates requires some prescription
of the probability of various configuration. In the absence of a physical
model supporting the distribution, one might attempt an effective
approach, where the fluctuation of the structure would be simulated
by fluctuations of the transition energies of the carotenoids (disorder
parameter $\Delta$) in several selected structures. If successful,
it could provide guidance in constructing a more detailed model of
aggregation which would take into account the thermodynamics of the
process. 

The exciton model is not able to account for the reduced extinction
coefficient of the aggregates. The total calculated extinction coefficient
of the aggregates would thus be overestimated. As a result, one cannot
expect to deduce actual concentrations of the aggregates from our
model, because the transition dipole moment of an aggregate is reduced
by an unknown factor. In order to fit the experimental spectra, we
will therefore calculate normalized absorption spectra $\alpha_{1}(\omega)$,
$\alpha_{2}(\omega,J_{2})$, $\alpha(\omega,J_{3})$,$\dots$ for
monomer, dimer, trimer and larger aggregates with some values $J$
of the resonance coupling. We will then combine them to get the total
normalized absorption spectrum $\alpha(\omega)$ as 
\begin{equation}
\alpha(\omega)=\frac{1}{{\cal N}}\sum_{n=1}^{K}a_{k}\alpha_{k}(\omega,J_{k}),\label{eq:abs_fit}
\end{equation}
where ${\cal N}$ is a normalization constant. This will give us an
qualitative picture of the ratio of various aggregates in the mixture,

The values of $J$ and the selected structures where chosen according
to the features observed in the astaxanthin spectra at 50 \% and 90
\% of water (see Fig.~\ref{fig:TPo1}A) while taking into account
qualitative results of the molecular orbital simulations. Such a procedure
cannot lead to unique results, but still seriously constrains the
composition of the aggregate spectra, as demonstrated below.

\section{Discussion}

\subsection{J-Aggregate Spectrum}

\begin{figure}[h]
\includegraphics[width=1\columnwidth]{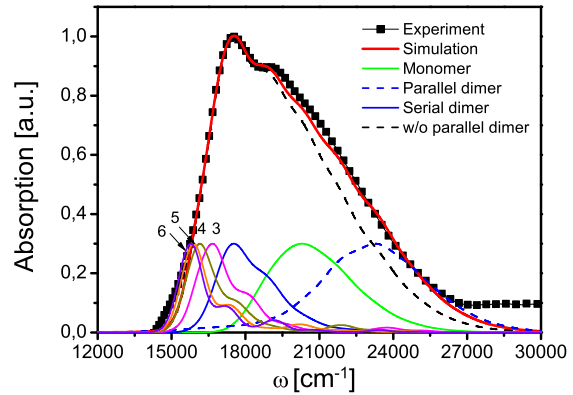}

\caption{\label{fig:A50procf}Absorption spectrum of a mixture of astaxanthin
aggregates at 50 \% of water. Theoretical spectrum is composed of
aggregates up to six members with $J<0$ (parallel dimer) and of a
sizable amount of parallel dimers with $J>0$ (see Tab.~\ref{tab:Parameters-50}).
The spectra of various aggregates are all normalized for their maximum
equal to 0.3.}
\end{figure}

Let us first analyze the spectra measured at $50$ \% of water. The
experimental spectrum shows a significant red shift by $\approx2800$
cm$^{-1}$. This is consistent with formation of J-aggregates ($J<0$)
i.e. with the ``serial'' dimer and larger aggregates of the similar
form. Combination of the monomeric spectrum with a spectrum of the
dimer ($J=-2500$ cm$^{-1}$) leads to the spectrum with characteristic
shoulder. Adding small portions of trimers, tetramers and larger aggregates
up-to six members leads to a good fit of the experimental spectrum
(black dashed curve in Fig.~\ref{fig:A50procf}) starting from $16000$
cm$^{-1}$, with slightly insufficient absorption for frequencies
larger than $20000$ cm$^{-1}$. The spectral region between $14000$
cm$^{-1}$ and $16000$ $^{-1}$ probably contains aggregates with
even more members than six. By adding the parallel dimer spectrum
(taken from the fit of the spectra at 90 \% of water, see below) we
obtain far better fit for the frequencies above $20000$ cm$^{-1}$
confirming the presence of some parallel H-aggregate type dimers already
at $50$ \% of water. 

The parameters of the fit are summarized in Tab.~\ref{tab:Parameters-50}.
The fit is composed of absorption spectra of small aggregates whose
constituents have the parameters same as the monomer in Fig.~\ref{fig:Monomer}
. The influence of the trimer and larger aggregate spectra on the
total spectrum is small and it was not possible to make conclusions
on the magnitude of the disorder in them. We leave their disorder
the same as in the monomer. 

\begin{table}[h]
\begin{tabular}{|c|c|c|c|}
\hline 
Nr. of members & $\Delta$ {[}cm$^{-1}${]} & $J$ {[}cm$^{-1}${]} & $\frac{a_{n}}{{\cal N}}$\tabularnewline
\hline 
\hline 
1 & 800 & - & 0.50\tabularnewline
\hline 
2 & 800 & -2500 & 0.75\tabularnewline
\hline 
3 & 800 & -2500 & 0.28\tabularnewline
\hline 
4 & 800 & -2500 & 0.09\tabularnewline
\hline 
5 & 800 & -2500 & 0.03\tabularnewline
\hline 
6 & 800 & -2500 & 0.02\tabularnewline
\hline 
2 & 2400 & 2100 & 0.21\tabularnewline
\hline 
\end{tabular}

\caption{\label{tab:Parameters-50}Parameters of the fit in Fig.~\ref{fig:A50procf}.
First column represents the number of monomers in an aggregate. The
transition energies $\hbar\omega_{eg}$ are those of the monomer in
Fig.~\ref{fig:Monomer} (parameters in the text) with disorder parameter
$\Delta$ and nearest neighbor coupling $J$. The sum spectrum is
composed of the normalized spectra of various aggregates according
to Eq.~(\ref{eq:abs_fit}) with coefficients $a_{n}$. The last row
is the parallel dimer from Tab.~\ref{tab:Parameters-90}.}
\end{table}

\subsection{H-Aggregate Spectrum}

\begin{figure}[h]
\includegraphics[width=1\columnwidth]{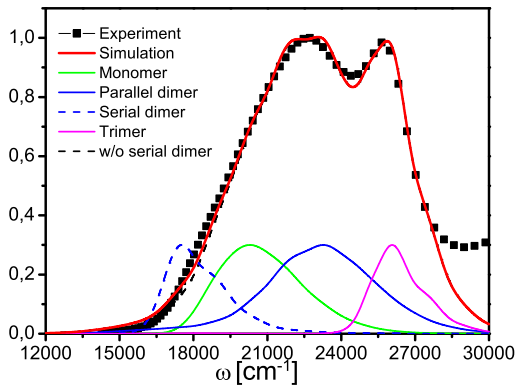}

\caption{\label{fig:A90procf}Absorption spectrum of a mixture of astaxanthin
aggregates at 90 \% of water. The spectrum is composed of mono-, di-
and trimers with $J>0$ (parallel dimer) and of a small amount of
``serial'' dimers with $J<0$ (see Tab.~\ref{tab:Parameters-90}).
The spectra of various aggregates are all normalized for their maximum
equal to 0.3.}
\end{figure}

At the concentration of water equal to 90 \%, we can assume the formation
of parallel dimers and possibly larger aggregates of similar structure.
Positive resonance coupling is suggested by the blue shift of the
spectrum with respect to the spectrum of the monomer. The line shape
can be fitted by a mixture of monomers, dimers and trimers of the
parallel type, and a small contribution of the ``serial'' dimers.
The parallel dimer has to have a significantly larger disorder than
the monomer, suggesting the existence of coupling disorder or some
deformation of the chain . The trimer contribution cannot be explained
by simple stacking of the monomers. This would result in a nearest
neighbor coupling and a second nearest neighbor coupling significantly
reduced, and correspondingly in an insufficient shift of the trimeric
peak. We have to rather assume an increased excitonic coupling which
acts among all three molecules of the aggregate. This is consistent
with all three molecules touching by their conjugated chains (see
Fig.~\ref{fig:Characteristic-structures-of}D) and a compactification
of the aggregate (molecules are closer to each other than in the dimer).
As an alternative to this fit, one could try to explain the most blue
peak of the experimental H-aggregate spectrum of astaxanthin (Fig.
\ref{fig:A90procf}) by a dimer with large resonance coupling. This
coupling would, however, have to be almost an order of magnitude larger
than the one calculated by our semi-empirical method. Assuming a compactified
trimer returns the required values of resonance coupling into a reasonable
parameter region.

\begin{table}[h]
\begin{tabular}{|c|c|c|c|}
\hline 
Nr. of members & $\Delta$ {[}cm$^{-1}${]} & $J$ {[}cm$^{-1}${]} & $\frac{a_{n}}{{\cal N}}$\tabularnewline
\hline 
\hline 
1 & 800 & - & 0.36\tabularnewline
\hline 
2 & 2400 & 2100 & 0.87\tabularnewline
\hline 
3 & 1200 & 2750 & 0.60\tabularnewline
\hline 
2 & 800 & -2500 & 0.06\tabularnewline
\hline 
\end{tabular}

\caption{\label{tab:Parameters-90}Parameters of the fit in Fig.~\ref{tab:Parameters-90}
First column represents the number of monomers in an aggregate. The
transition energies $\hbar\omega_{eg}$ are those of the monomer in
Fig.~\ref{fig:Monomer} (parameters in the text) with disorder parameter
$\Delta$. In case of a dimer, $J$ represents the coupling between
the two monomers, for trimer $J$ is the coupling of a monomer to
other two monomers. The sum spectrum is composed of the normalized
spectra of parallel mono-, di- and trimers and ``serial'' dimer.
The sum spectrum is composed of the normalized aggregate spectra according
to Eq.~(\ref{eq:abs_fit}) with coefficients $a_{n}$. The last row
is the ``serial'' dimer from Tab.~\ref{tab:Parameters-50}.}
\end{table}

In both investigated absorption spectra (Figs.~\ref{fig:A50procf}
and \ref{fig:A90procf}) we can see several distinct excitonic features.
The most significant feature is the spectral shift of the bands. However,
the line shapes also bear signatures of band narrowing, evident e.g
in the appearance of the vibrational shoulder in the dimer spectrum
at $50$ \% of water, or in the narrow line shape of the trimers at
$90$ \% of water. Both these shapes determine the most obvious features
of the aggregate spectra.

\section{Conclusions}

We have measured and analyzed absorption spectra of aggregates of
the carotenoid astaxanthin in hydrated dimethylsulfoxide. We have
demonstrated that these absorption spectra can be explained quantitatively
as a sum of spectra of small aggregates. Phenomenological excitonic
model of resonance interaction between $\pi-$conjugated chains of
astaxanthin molecules was constrained by semi-empirical calculations.
The observed spectral shifts were assigned to formation of J-type
and H-type aggregates and their changing ratio under different concentration
of water. Fitting experimental absorption spectra lead to a qualitative
estimation of ratio of aggregates of various types in the sample at
50 \% and 90 \% of water in dimethylsolfoxide. With increased percentage
of water in the solvent, the proportion of compact aggregates with
a decreasing surface area exposed to the solvent is growing. Based
on absorption spectra, it is not possible to exclude that different
aggregate configurations result in the same observed features. The
ease with which the spectra can be explained by the spectra of small
aggregates, however, provides a relatively high degree of confidence
in our results.
\begin{acknowledgments}
The research was supported by grants from the Czech Ministry of Education
(MSM6007665808) and the Czech Science Foundation (202/09/1330). J.O.
acknowledges the support of GAUK through grant nr. 416311. The authors
thank Lucie T\v{e}snohl\'{i}dkov\'{a} for her help with preparation of astaxanthin
aggregates.
\end{acknowledgments}
\bibliographystyle{prsty}
\bibliography{carsDMSO}

\begin{thebibliography}{10}

\bibitem{Polivka2004a-1}
T. Pol\'{i}vka and V. Sundstr\"{o}m, Chem. Rev. {\bf 104},  2021  (2004).

\bibitem{Ostroumov-2}
E. Ostroumov, M. Muller, C. Marian, M. Kleinschmidt, and A. Holzwarth, Phys.
  Rev. Lett. {\bf 103},  108302  (2009).

\bibitem{Maiuri-3}
M. Maiuri, D. Polli, D. Brida, L. Luer, A. LaFountain, M. Fuciman, R.~J.
  Cogdell, H. Frank, and G. Cerullo, Phys. Chem. Chem. Phys. {\bf 14},  6312
  (2012).

\bibitem{Polivka-4}
T. Pol\'{i}vka and V. Sundstr\"{o}m, Chem. Phys. Lett. {\bf 477},  1  (2009).

\bibitem{Kosumi-5}
D. Kosumi, T. Kusumoto, R. Fujii, M. Sugisaki, Y. Iinuma, N. Oka, Y. Takaesu,
  T. Taira, M. Iha, H.~A. Frank, and H. Hashimoto, Phys. Chem. Chem. Phys. {\bf
  22},  10762  (2011).

\bibitem{Gradinaru-6}
C.~C. Gradinaru, J. Kennis, E. Papagiannakis, I. van Stokkum, R. Cogdell, G.
  Fleming, R. Niederman, and R. van Grondelle, Proc. Natl. Acad. Sci. U.S.A.
  {\bf 98},  2364  (2001).

\bibitem{Polivka-7}
T. Pol\'{i}vka and H.~A. Frank, Acc. Chem. Res. {\bf 43},  1125  (2010).

\bibitem{Ruban-8}
A. Ruban, R. Berera, C. Ilioaia, van Stokkum~I.H.M., J. Kennis, A. Pascal, H.
  van Amerongen, B. Robert, P. Horton, and R. van Grondelle, Nature {\bf 450},
  575  (2007).

\bibitem{Koyama-9}
Y. Koyama, F.~S. Rondonuwu, R. Fujii, and Y. Watanabe, Biopolymers {\bf 74},  2
   (2004).

\bibitem{Starcke-10}
J.~H. Starcke, M. Wormit, J. Schirmer, and A. Dreuw, Chem. Phys. {\bf 329},  39
   (2006).

\bibitem{Ghosh-11}
D. Ghosh, J. Hachmann, T. Yanai, and G.~K.~L. Chan, J. Chem. Phys. {\bf 128},
  144117  (2008).

\bibitem{Kleinschmidt-12}
M. Kleinschmidt, M.~M. C, M. Waletzke, and S. Grimme, J. Chem. Phys. {\bf 130},
   044708  (2009).

\bibitem{Simonyi-13}
M. Simonyi, Z. Bikadi, F. Zsila, and J. Deli., Chirality {\bf 15},  680
  (2003).

\bibitem{Billsten-14}
H.~H. Billsten, V. Sundstr\"{o}m, and T. Pol\'{i}vka, J. Phys. Chem. A {\bf 109},  1521
   (2005).

\bibitem{Koepsel-15}
C. K\"{o}psel, H. M\"{o}ltgen, H. Schuch, H. Auweter, K. Kleinermanns, H.-D. Martin,
  and H. Bettermann, J. Mol. Struct. {\bf 750},  109  (2005).

\bibitem{Mori-16}
Y. Mori, K. Yamano, and H. Hashimoto, Chem. Phys. Lett. {\bf 254},  84  (1996).

\bibitem{Ruban-17}
A.~V. Ruban, P. Horton, and A.~J. Young, J. Photochem. Photobiol. B {\bf 21},
  229  (1993).

\bibitem{Giovannetti-18}
R. Giovannetti, Alibabaei, L. Pucciarelli, and F., Spectrochim. Acta A {\bf
  73},  157  (2009).

\bibitem{Spano-19}
F. Spano, J. Am. Chem. Soc. {\bf 131},  4267  (2009).

\bibitem{Wang-20}
C. Wang and T. M.J., J. Am. Chem. Soc. {\bf 132},  13988  (2010).

\bibitem{Wang-21}
C. Wang, D.~E. Schlamadinger, V. Desai, and M.~J. Tauber, Chem. Phys. Chem.
  {\bf 12},  2891  (2011).

\bibitem{Sujak-22}
A. Sujak, P. Mazurek, and W.~I. Gruszecki, J Photochem. Photobiol. B {\bf 68},
  39  (2002).

\bibitem{Gruszecki-23}
W. Gruszecki,  in {\em Photochemistry of Carotenoids}, edited by G.~B.
  H.~A.~Frank, A. J.~Young and R.~J. Cogdell (Kluwer Academic Publishers,
  Dordrecht, Netherlands, 1999), Chap.~Carotenoids in membranes, p.\ 363.

\bibitem{Aspinall-24}
M. Aspinall-O'Dea, M. Wentworth, A. Pascal, B. Robert, A. Ruban, and P. Horton,
  Proc. Natl. Acad. Sci. U.S.A. {\bf 99},  16331  (2002).

\bibitem{chabera-25}
P. Ch\'{a}bera, M. Durchan, P.~M. Shih, C.~A. Kerfeld, and T. Pol\'{i}vka,
  BBA-Bioenergetics {\bf 1807},  30  (2011).

\bibitem{Gao-26}
F.~G. Gao, A.~J. Bard, and L.~D. Kispert, J. Photochem. Photobiol. A {\bf 130},
   49  (2000).

\bibitem{Pan-27}
J. Pan, G. Benk\"{o}, Y.~H. Xu, T. Pascher, L. Sun, V. Sundstr\"{o}m, and T. Pol\'{i}vka,
  J. Am. Chem. Soc. {\bf 124},  13949  (2002).

\bibitem{Pan-28}
J. Pan, Y. Xu, L. Sun, V. Sundstr\"{o}m, and T. Pol\'{i}vka, J. Am. Chem. Soc. {\bf
  126},  3066  (2004).

\bibitem{Xiang-29}
J. Xiang, F.~S. Rondonuwu, Y. Kakitani, R. Fujii, Y. Watanabe, Y. Koyama, H.
  Nagae, Y. Yamano, and M. Ito, J. Phys. Chem. B {\bf 109},  17066  (2005).

\bibitem{Wang-30}
X.~F. Wang, Y. Koyama, H. Nagae, Y. Yamano, M. Ito, and Y. Wada, Chem. Phys.
  Lett. {\bf 420},  309  (2006).

\bibitem{Smith-31}
J. Smith, M.B.and~Michl, Chem. Rev. {\bf 110},  6891  (2010).

\bibitem{Cianci-32}
M. Cianci, R. P.J., A. Olczak, J. Raftery, N. Chayen, P. Zagalsky, and J.
  Helliwell, Proc. Natl. Acad. Sci. U.S.A. {\bf 15},  9795  (2002).

\bibitem{Wijk-33}
A. van Wijk, A. Spaans, N. Uzunbajakava, C. Otto, H. De~Groot, J. Lugtenburg,
  and F. Buda, J. Am. Chem. Soc. {\bf 127},  1438  (2005).

\bibitem{Frank-34}
F.~H. A., J.~A. Bautista, J. Josue, Z. Pendon, R.~G. Hiller, F.~P. Sharples, D.
  Gosztola, and M.~R. Wasielewski, J Phys Chem B {\bf 104},  4569  (2000).

\bibitem{Zigmantas-35}
D. Zigmantas, R.~G. Hiller, F.~P. Sharples, H.~A. Frank, V. Sundstr\"{o}m, and
  T. Polivka, Phys Chem Chem Phys {\bf 6},  3009  (2004).

\bibitem{Doll2008a}
R. Doll, D. Zueco, M. Wubs, S. Kohler, and P. Hanggi, Chem. Phys. {\bf 347},
  243  (2008).

\bibitem{Cho2005a}
M.~H. Cho, H.~M. Vaswani, T. Brixner, J. Stenger, and G.~R. Fleming, J. Phys.
  Chem. B {\bf 109},  10542  (2005).

\bibitem{Wang2012}
C. Wang, C.~J. Berg, C.-C. Hsu, B.~A. Merril, and M.~J. Tauber, J. Phys. Chem.
  B  article ASAP  (2012).

\bibitem{MukamelBook}
S. Mukamel, {\em Principles of nonlinear spectroscopy} (Oxford University
  Press, Oxford, 1995).

\bibitem{ValkunasBook}
H. van Amerongen, L. Valkunas, and R. van Grondelle, {\em Photosynthetic
  Excitons} (World Scientific, Singapore, 2000).

\bibitem{Wang-36}
J. Wang, R.~M. Wolf, J.~W. Caldwell, P.~A. Kollman, and D.~A. Case, J. Comput.
  Chem. {\bf 25},  1157  (2004).

\bibitem{Berendsen-37}
H.~J.~C. Berendsen, J.~R. Grigera, and T.~P. Straatsma, J. Phys. Chem. {\bf
  91},  6269  (1987).

\bibitem{gaussian03-38}
M.~J. Frisch, G.~W. Trucks, H.~B. Schlegel, G.~E. Scuseria, M.~A. Robb, J.~R.
  Cheeseman, J. Montgomery, J.~A., T. Vreven, K.~N. Kudin, J.~C. Burant, J.~M.
  Millam, S.~S. Iyengar, J. Tomasi, V. Barone, B. Mennucci, M. Cossi, G.
  Scalmani, N. Rega, G.~A. Petersson, H. Nakatsuji, M. Hada, M. Ehara, K.
  Toyota, R. Fukuda, J. Hasegawa, M. Ishida, T. Nakajima, Y. Honda, O. Kitao,
  H. Nakai, M. Klene, X. Li, J.~E. Knox, H.~P. Hratchian, J.~B. Cross, V.
  Bakken, C. Adamo, J. Jaramillo, R. Gomperts, R.~E. Stratmann, O. Yazyev,
  A.~J. Austin, R. Cammi, C. Pomelli, J.~W. Ochterski, P.~Y. Ayala, K.
  Morokuma, G.~A. Voth, P. Salvador, J.~J. Dannenberg, V.~G. Zakrzewski, S.
  Dapprich, A.~D. Daniels, M.~C. Strain, O. Farkas, D.~K. Malick, A.~D. Rabuck,
  K. Raghavachari, J.~B. Foresman, J.~V. Ortiz, Q. Cui, A.~G. Baboul, S.
  Clifford, J. Cioslowski, B.~B. Stefanov, G. Liu, A. Liashenko, P. Piskorz, I.
  Komaromi, R.~L. Martin, D.~J. Fox, T. Keith, M.~A. Al-Laham, C.~Y. Peng, A.
  Nanayakkara, M. Challacombe, P.~M.~W. Gill, B. Johnson, W. Chen, M.~W. Wong,
  C. Gonzalez, and J.~A. Pople, Gaussian, Inc., Wallingford CT, 2004.

\bibitem{Wang-39}
J. Wang, W. Wang, P.~A. Kollman, and D.~A. Case, J. Mol. Graph. Model {\bf 25},
   247260.  (2006).

\bibitem{amber8-40}
D.~A. Case, T.~A. Darden, T.~E.~I. Cheatham, C.~L. Simmerling, J. Wang, R.~E.
  Duke, R. Luo, K.~M. Merz, B. Wang, D.~A. Pearlman, M. Crowley, S. Brozell, V.
  Tsui, H. Gohlke, J. Mongan, V. Hornak, G. Cui, P. Beroza, C. Schafmeister,
  J.~W. Caldwell, W.~R. Ross, and P.~A. Kollman, Amber 8, University of
  California, San Francisco, 2004.

\bibitem{Berendsen-41}
H.~J.~C. Berendsen, J.~P.~M. Postma, W.~F. van Gunsteren, A. DiNola, and J.~R.
  Haak, J. Chem. Phys. {\bf 81},  3684  (1984).

\bibitem{Daren-42}
T. Darden, D. York, and L.~G. Pedersen, J. Chem. Phys. {\bf 98},  10089
  (1993).

\bibitem{Essmann-43}
U. Essmann, L. Perera, M.~L. Berkowitz, T. Darden, H. Lee, and L.~G. Pedersen,
  J. Chem. Phys. {\bf 103},  8577  (1995).

\bibitem{Ryckaert-44}
J.~P. Ryckaert, G. Ciccotti, and H.~J.~C. Berendsen, J. Comput. Phys. {\bf 23},
   327  (1977).

\bibitem{Humphrey-45}
W. Humphrey, A. Dalke, and K. Schulten, J. Mol. Graphics {\bf 14},  33  (1996).

\bibitem{Szabo1996a}
A. Szabo and N.~S. Ostlund, {\em Modern Quantum Chemistry Introduction to
  Advanced Electronic Structure Theory} (Dover, New York, 1996).

\bibitem{Hoffmann1963}
R. Hoffmann, J. Chem. Phys. {\bf 39},  1397  (1963).

\bibitem{Smith2004}
S.~M. Smith, A.~N. Markevitch, D.~A. Romanov, X. Li, R.~J. Levis, and H.~B.
  Schlegel, J. Phys. Chem. A {\bf 108},  11063  (2004).

\bibitem{Madjet2006}
M.~E. Madjet, A. Abdurahman, and T. Renger, J. Phys. Chem. B {\bf 110},  17268
  (2006).

\end{thebibliography}

\end{document}